# Using perceptually defined music features in music information retrieval

Anders Friberg[1], Erwin Schoonderwaldt[2], Anton Hedblad[1], Marco Fabiani[1], and Anders Elowsson[1]

[1]KTH Royal Institute of Technology

School of Computer Science and Communication, Speech, Music and Hearing

Stockholm, Sweden

[2]Hanover University of Music, Drama and Media

Institute of Music Physiology and Musicians' Medicine

Hannover, Germany

Running title: perceptual music features

**Abstract**

In this study, the notion of *perceptual features* is introduced for describing general music properties based on human perception. This is an attempt at rethinking the concept of features, in order to understand the underlying human perception mechanisms. Instead of using concepts from music theory such as tones, pitches, and chords, a set of nine features describing overall properties of the music was selected. They were chosen from qualitative measures used in psychology studies and motivated from an ecological approach. The selected perceptual features were rated in two listening experiments using two different data sets. They were modeled both from symbolic (MIDI) and audio data using different sets of computational features. Ratings of emotional expression were predicted using the perceptual features. The results indicate that (1) at least some of the perceptual features are reliable estimates; (2) emotion ratings could be predicted by a small combination of perceptual features with an explained variance up to 90%; (3) the perceptual features could only to a limited extent be modeled using existing audio features. The results also clearly indicated that a small number of dedicated features were superior to a "brute force" model using a large number of general audio features.

# I. INTRODUCTION

One of the fundamental research goals in music information retrieval (MIR) research is to perform a content-based analysis of music represented in audio files. This has resulted in a large number of computational features suggested in the literature (Burred and Lerch, 2004; Peeters, 2004; Polotti and Rocchesso, 2008). The features can be broadly divided into two categories. The first is low-level features, often based on short-time, frame-based measures. They consist typically of different spectral features such as MFCC coefficients, spectral centroid, or the number of zero crossings per time unit, but also psychoacoustic measures such as roughness and loudness models. The second is mid-level features with a longer analysis window. These features are typically well-known concepts from music theory and music perception such as beat strength, rhythmic regularity, meter, mode, harmony, and key. They are often verified by using ground-truth data with examples annotated by experts. In addition, a third level consists of semantic descriptions such as emotional expression or genre. However, the distinction between the different levels is in reality often rather vague and was made in order to point to some general differences in complexity and aims.

A starting point in computational analysis of music is often basic music theory, which has a long tradition of analyzing the music into its components. These components are perceptual aspects like functional harmony or rhythmic patterns. However, the perspective is often that of an ideal expert listener who is able to perceive the music in extreme detail including all played notes. By contrast, in a natural listening situation it is not possible to hear such depth of detail even after extensive practice. This indicates that other, coarser perceptual features are perceived in a realistic listening setting. If we aim to model higher-level concepts such as emotion description or genre, it is unlikely that the mid-level (or low-level) features derived from classical music theory (or low-level signal properties) are the best choices. In fact, in emotion research, a number of rather imprecise overall estimations have long been successfully used. Examples are overall pitch height, overall dynamics or harmonic complexity (e.g. Gabrielsson and Lindström, 2010; Juslin and Lindström 2010; Eerola, Friberg and Bresin, 2013; Friberg, 2008). This indicates that human music perception retrieves information that might be different from traditional concepts of music theory, such as the harmonic progression. This is not surprising since it demands substantial training to recognize an harmonic progression, but it also points to the need for finding what we really hear when we listen to music.

Another alternative is to view music perception from the ecological perspective (Clarke, 2005). A general principle in ecological perception is that humans always try to understand the world from the sound. This means that we try to understand the source properties of the sounds rather than only considering the specific timbre quality, for example. This is evident for everyday sounds. Take, for example, a cutlery sound in the kitchen. It is not heard as a certain "cling" sound but the sound of the specific objects (Gaver, 1993). Thus, according to the ecological hypothesis, it would be easier (or more natural) for humans to perceive and estimate the speed of the music by relating it to movement rather than the event rate or to perceive the energy used to produce the sound rather than the sound level (Ladefoged and McKinney, 1963).

In this paper, we will present a new method for determining audio features based on perceptual ratings and discuss their potential in MIR-based audio analysis systems. We will call these *perceptual features* to emphasize that they are by definition based on perception even though they are estimated computationally. An outline of the proposed analysis model is shown in Figure 1.

*** Figure 1 about here ***

Why did we choose to work with perceptual features? The reason was mainly to acquire a deeper understanding of what we really hear when we listen to music. Is there some intermediate analysis layer that is processed and in that case what does it consists of? These are very difficult questions to address and we do not claim that the perceptual features proposed here form the answer. They do, however, represent an attempt to start "afresh" using a broader view on perception, looking from different angles and considering different disciplines of research.

Another reason for working with perceptual features was the current focus on, and use of, features in different computational models within the MIR community. The prevailing focus is often on constructing models that can explain as much of the target variation as possible, thus a pure engineering approach. This is manifested in the goal of the MIREX competition (Downie, West, Ehmann, and Vincent. 2005). In order to achieve better results, more and more features are devised and tested, often based on intuition and trial-and-error. This has resulted in a very large number of proposed features in the literature and a current approach is to also automatically derive the features (e.g. Pachet and Roy, 2009). Previous extensive testing of different feature sets has indicated that it is hard to reach beyond a certain performance, thus suggesting that there is actually a theoretical limit, a "glass ceiling" when using this paradigm (Aucouturier and Pachet, 2004; Pohle, Pampalk and Widmer, 2005).

Although many of the proposed audio features are of a similar nature, it is unlikely that human perception uses all of them. As also discussed above, previous research in emotion communication has shown that less than a handful of qualitative features can explain a major part of the listener responses. In the experiments both by Juslin and Lindström (2010) and by Eerola Friberg and Bresin (2013), the explained variance amounted to nearly 90%, which is mostly above what has been obtained in previous audio analysis methods. In these studies, the feature variation was obtained by synthesizing different music examples. One of the most successful models using audio features was made by Lu, Liu, and Zhang (2006): it obtained an average recognition of 86% using a few custom features applied on a database with classical music. This indicates that the suggestion of a glass ceiling here is wrong (or a little higher) and that it is possible to reach a better performance, as in this case, by using a restricted set of dedicated features.

Most MIR research is based on human perception since the top-level ground-truth data mostly consists of human annotations. The idea in the present model is to also use perceptually derived data one level down in the feature space. It will still be a challenge to model these perceptual features from audio. However, the advantage is that with a careful selection of perceptual features, the scope of each feature is more limited to certain kinds of parameters, such as low-level features related to onset detection, and thus potentially easier to model than a higher level feature model. Such models are also straightforward to assess since they rely directly on perceptual data and thus can be evaluated in the same way as they were determined in the present study. A perceptual/neural approach has been used before, for example, in models of the nerve responses in the cochlea, by using the auditory model by Slaney (1998), for instance. This method is difficult to extend to mid-level features since knowledge about higher level processing in the human hearing system along the auditory pathway is still scarce, although recently, work has been carried out on speech using spectro-temporal fields mimicking properties of higher-level auditory neurons (Mesgarani, Shamma and Slaney, 2004; Heckmann, Domont, Joublin and Goerick, 2011).

An interesting discussion on the direction of MIR research and methods in relation to cognitive psychology has recently been presented by Aucouturier and Bigand (2013). They suggest that one of the weaknesses of the current MIR direction is the lack of psychological validation for low-level features. The current approach addresses this problem, however, not by providing validation of the used features but rather by estimating the features directly, using perceptual experiments.

We will check the usability of perceptual features in MIR applications by asking the following questions:

*-Do the raters agree in their judgments?*

*-How many raters do we need?*

*-Can higher-level semantic descriptions be accurately modeled from perceptual features?*

*-Can we make computational models of the perceptual features?*

In this paper, we will not give an exhaustive answer to each question but present the current state and experimental data that has been collected so far within this ongoing project. In an initial study (Friberg, Schoonderwaldt and Hedblad, 2011), we examined and motivated the selection of perceptual features, made a pilot perceptual experiment to select a set of music examples consisting of polyphonic ring tones, and made a final estimation of the perceptual features using this reduced set. In another study (Hedblad, 2011; Friberg and Hedblad, 2011), a preliminary estimation of the perceptual features was made from acoustical features on the ring tone data set. In the current article, which attempts to summarize the results, the previous work is extended with a second database (film clips) for all analyses, a new analysis of MIDI features, and with a new estimation of the perceptual features from audio features using alternative methods and cross-validation.

Before going into the questions stated above, we will present the databases used, the selected features and the perceptual rating experiments.

## II. METHOD

### A. Music data

Two different music collections were used. The *ring tone data* consists of 100 polyphonic ring tones that were converted to audio. Initially, a set of 242 ringtones was randomly picked from a large commercial database consisting of popular music of various styles encoded in MIDI symbolic representation. The final set of 100 ring tones was selected by optimizing the spread of each feature based on the result from a pilot experiment (Friberg, Schoonderwaldt and Hedblad, 2011). In the majority of cases, they were instrumental polyphonic versions of the original popular songs. The average duration of the ringtones was about 30 s. The MIDI files were converted to audio using a Roland JV-1010 MIDI synthesizer. The resulting uncompressed wave files (16-bit resolution, 44.1 kHz sampling frequency) were normalized according to the loudness standard specification ITU-R BS. 1770 (ITU-R, 2006) using the implementation by Nygren (2009).

The *film clips* consist of 110 short audio excerpts from film music. This set was provided by the University of Jyväskylä and has been used in different experiments on emotional communication in music (e.g. Eerola and Vuoskoski, 2011). The examples had been carefully selected to exhibit a variation in different emotional expressions and were previously rated according to a variety of emotional expressions including both discrete and dimensional emotion models. In this study, the film clips were not normalized in the same way as the ring tones since that would potentially have distorted the ratings of the emotional expressions.

**B. Perceptual features**

The selection of perceptual features was motivated by their relevance in emotion research and by using the ecological perspective, as discussed above. Several of these features were used by Wedin (1972) in a similar experiment. Due to experimental time constraints, the final number of features was limited to nine basic feature scales and two emotion scales. See Friberg, Schoonderwaldt and Hedblad, (2011) for a further discussion of the selection process. For each feature below, the extremes of the rating scale are indicated in parentheses.

*Speed (slow-fast)* The overall speed of the music. We wanted to avoid the more complicated notion of tempo and use something that is easy for both musicians and non-musicians to relate to; see also Madison and Paulin (2010).

*Rhythmic clarity (flowing-firm)* The amount of rhythmic accentuation disregarding the rhythm pattern. This would presumably be similar to pulse clarity as modeled by Lartillot, Eerola, Toiviainen and Fornari (2008).

*Rhythmic complexity (simple-complex)* This is a natural companion to rhythmic clarity and relates more to the differences in rhythmic patterns.

*Articulation (staccato-legato)* The overall articulation related to the duration of tones in terms of *staccato* or *legato*.

*Dynamics (soft-loud)* The overall dynamic level as obtained from timbre and other cues presumably disregarding listening volume. Note that the stimuli were normalized using an equal loudness model. It is presumably related to the estimated effort of the player.

*Modality (minor-major)* Contrary to music theory, modality is rated here on a continuous scale ranging from minor to major.

*Overall Pitch (low-high)* The overall pitch height of the music.

*Harmonic complexity (simple-complex)* A measure of how complex the harmonic progression is. It might reflect, for example, the amount of chord changes and deviations from a certain key scale structure. This is presumably a difficult feature to rate, demanding some knowledge of music theory.

*Brightness (dark-bright) (Exp. 1)*, *Timbre (Exp. 2)* One general timbre parameter. Two different terms were used since the results from Exp. 1 indicated some misunderstanding by the raters.

In addition, the following emotional scales were rated in Exp. 1. For the purpose of this experiment, they were regarded as high-level descriptions and as dependent variables in the subsequent analysis.

*Energy (low-high) (Exp. 1)* The overall perceived energy. This is similar to (but not the same as) the terms Activity or Arousal in the common two-dimensional emotion model (e.g. Russell, 1980). As for dynamics above, it is also presumably an important ecological aspect for estimating the energy of the source.

*Valence (negative-positive) (Exp. 1)* Indicates the degree of negative or positive emotion.

For the film clips in Exp. 2 there were several emotion ratings available from Eerola and Vuoskoski (2011). In this study, we used the previous ratings for Energy, Valence, Tension, Anger, Fear, Happiness, Sadness, and Tenderness.

**C. Rating experiments**

Two listening experiments were conducted with 20 and 21 subjects, respectively. All subjects had some limited musical experience usually at an amateur level and most of them were students at KTH. The average age in Exp. 1 was 30 years (range 18-55; 7 women, 13 men). They listened to music on average 15 hours a week (range 3-40) and played at least one instrument. The average number of years playing their main instrument was 14 years (range 3-45). The average age in Exp. 2 was 32 years (range 21-56; 4 women, 18 men). They listened to music on average 15 hours a week (range 2-45) and played at least one instrument. The average number of years playing their main instrument was 16 years (range 1-32).

The procedure and experimental setup was the same for both experiments. The stimuli were presented using two studio loudspeakers (Genelec 1031A) at a distance of about 2 m from the listener in a quiet room. The sound level was calibrated to 75 dB(A) at the approximate listening position using low pass filtered noise roughly corresponding to the spectral content of the stimuli. Each subject individually rated all features and in the first experiment the emotion descriptions on quasi-continuous (9 steps) Likert scales for each music example were presented on a computer screen. All examples were randomized for each subject. The subjects were free to take a break at any time and the whole test took 1-2.5 hours. They were each reimbursed with two cinema tickets. Further details of the experimental setup in Exp. 1 are given in Friberg, Schoonderwaldt and Hedblad, (2011).

## III. RESULTS

### A. Listener agreement

The listeners' agreement in terms of the individual variation was estimated by the mean Pearson's correlation *r* between all subject pairs; see Table 1. The purpose in this study was to make the best possible estimate of the true mean of each perceptual feature. Thus, for estimating the reliability of the mean estimate, Cronbach's alpha was used. It is the same as the intra-class correlation ICC(C,k), case 2 (McGraw and Wong, 1996). The Cronbach's alpha indicated good agreement for all ratings. Commonly, a value higher than 0.9 is used to indicate excellent agreement of the data. This was the case for most features with the exception of harmonic complexity. However, the inter-subject correlations showed a larger variation across the features where the more complex features like harmonic complexity obtained a lower inter-subject correlation. The inter-subject correlations were useful for showing the actual variation of the responses, which was confirmed by informally observing the plotted variation.

A manual inspection of the inter-subject correlations revealed that there were some features in which one subject clearly deviated from the rest of the group. It was a different subject for each feature. For example, for rated modality one subject, who was omitted, misinterpreted the direction of the scale and answered in the opposite way to the average. Numbers in parentheses refer to the resulting data when these subjects were omitted. Since the purpose here was to obtain the best possible estimation of the mean values, this trimmed data was used in the subsequent analysis.

We interpret these results preliminarily as an indication that all the measures could tentatively be rated by the subjects. Although the more complex measures like rhythmic complexity obtained lower agreement, the mean value across subject may still be a fair estimate of the true mean in a larger population as indicated by the relatively high Cronbach's alphas.

*** Table 1 about here ***

**B. Independence of the rated features**

The independence of the rated features was estimated by computing the pair-wise correlation between all combinations of feature ratings; see 2. Ideally, some of the features, such as Speed and Dynamics, would be expected to be independently rated with little correlation. This was not the case here as indicated by the cross-correlations across features, but a rather modest correlation was obtained in a majority of cases. In the first experiment, about half of the correlations were significant and rarely exceeded 0.6 (corresponding to 36% covariation). The only exception was Pitch and Brightness with $r = 0.9$. A similar picture was obtained in Exp. 2 regarding the overall values. The reason for the cross-correlations in the ratings cannot be tracked down at this point since there are at least two different possibilities: either there is a covariation in the music examples, or alternatively, the listeners might not have been able to isolate each feature as intended. However, it seems obvious that there was a strong connection between Pitch and Brightness/Timbre in Exp. 1 and 2, as reflected by the subjects' rating. We acknowledged this dependency after performing Exp. 1 and subsequently tried to improve the experimental procedure using another term and by more careful instructions to the subjects. However, interestingly, this dependence also remained in Exp. 2, indicating that there might be a strong perceptual coupling between pitch and timbre.

*** Table 2 about here ***

# IV. MODELING HIGHER-LEVEL SEMANTIC DESCRIPTIONS FROM PERCEPTUAL FEATURES

A separate regression analysis was applied with each of the emotion ratings in both experiments as dependent variables and with all the nine perceptual features as independent variables. In order to check that an over-fitting did not occur, a 10-fold cross validation was also performed. Thus, the songs were randomly divided into 10 equal groups. For each group, the remaining 90% of the examples were used for training and the model was then evaluated on the group itself. This was further repeated for 50 different random selections and the result was averaged over the mean square error from each analysis. The results are summarized in Table 3. The overall amount of explained variance is estimated by $R^2$ or adjusted $R^2$. In addition, we used Partial Least-square Regression (PLS) and Support Vector Regression (SVR) as prediction methods (see the section below for the details). PLS and SVR almost never yielded better results than linear regression, and it often resulted in the same explained variance. Thus, we concluded that simple linear regression was adequate, indicating negligible effects of non-linearity and feature interaction in accordance with Juslin and Lindström (2010) and Eerola, Friberg and Bresin (2013).

In Exp. 1, the Energy rating could be predicted with an explained variance of 93% (adj. $R^2 = 0.93$) with four significant perceptual features (Speed, Rhythmic clarity, Dynamics, Modality). See Table 3. The strongest contribution was by Speed and Dynamics. The Valence rating in Exp. 1 was predicted with an explained variance of 87% (adj. $R^2 = 0.87$). The strongest contribution was by Modality followed by Dynamics (negative) and three other features (Speed, Articulation, Brightness). The fact that Energy was predicted mainly by a combination of Speed and Dynamics, and that Modality was the strongest predictor for Valence, seems very intuitive and corresponds to earlier experiments in emotion communication (e.g. Ilie and Thompson, 2006; Husain, Thompson and Schellenberg, 2002). The overall results were unexpectedly strong given the small number of perceptual features.

Note, however, that both the feature ratings and the emotion ratings were obtained from the same subjects, which might in theory explain some of the high correlations.

In Exp. 2, however, the emotion ratings were obtained in a separate experiment (Eerola and Vuoskoski, 2011), ensuring a higher independence between the rated features and rated emotions. Energy was still predicted with a high degree of explained variance (91% Adj. $R^2$) with Dynamics as the strongest predictor followed by Speed, Articulation and Modality (Table 3). Valence was rather well predicted with an explained variance of (78% Adj. $R^2$) mainly from Modality, Dynamics, and Harmonic complexity. The discrete emotions in Exp. 2 were rather well predicted for Happiness (81%), Sadness (75%), and Anger (72%), while Fear and Tenderness had the lowest prediction results (64% and 62%, respectively).

A comparison of the results for Energy and Valence in Exp. 1 and 2 indicates that the strongest predictors are the same in both experiments. There are some differences, possibly reflecting feature differences in the two databases. For example, the influence of Speed seems to be comparatively low in Exp. 2 across all emotions and there seems to be a focus on harmonic, tonal, timbral features instead. In particular, note that Harmonic complexity is a significant feature for most of the emotions in Exp. 2. This might reflect the most important features used by the composers of the film music examples, which intuitively seems to vary more in these respects.

Note that there are relatively few features, about four, that are significant for each emotion and still a fair amount of variation is explained. This supports the assumption that the selected features, at least in some cases, correspond to the most important features for emotional expression as suggested in previous studies (e.g. Gabrielsson and Lindström, 2010). This also shows that potentially, the number of computational features can be small given that they are modeled according to perceptual relevance and with high accuracy (See also Eerola, 2012).

\*\*\* Table 3 about here \*\*\*

# V. PREDICTION OF PERCEPTUAL FEATURES FROM SYMBOLIC DATA IN EXPERIMENT 1.

The ring tone database originally consisted of a collection of symbolic MIDI files. Thus, potentially, using the MIDI data directly would give rather accurate estimations of basic note parameters. Certainly, all notes are represented directly in the symbolic format and it is simple to compute different symbolic features such as note density or average pitch. However, what is lost in the MIDI representation are sonic properties of different notes and instruments, which are related to sound parameters in the synthesizer such as the timbre and sound level of different instruments.

The MIDI features were computed using a custom-made patch in the previously developed environment for music performance Director Musices (Friberg, Bresin and Sundberg, 2006). The tracks in the polyphonic MIDI files in a previous experiment were annotated as melody, accompaniment, bass or drums (Friberg and Ahlbäck, 2009). For each of these categories and for the whole score, four different features were computed. Only notes louder than 20 dB below the maximum sound level were used in all the computations. The main reason was to avoid some very soft notes in a few examples that could not be heard in the final mix.

*Note density* in terms of notes per second (NPS) was computed as the average number of notes across the whole example divided by the duration. Simultaneous onsets within a 50 ms time window were counted as one. This value was chosen as an intermediate point between the limit for melodic perception of about 80 ms (Friberg and Sundström, 2002; London, 2004) and the variation in ensemble playing with a standard deviation of about 20 ms (Rasch, 1979). For the drum track, a further division was made into two categories. NPS was computed separately for all tom sounds (toms, snare, bass drum) and the rest (hihat, cymbals).

*Sound level* (SL) was estimated as the average over all remaining notes after the softest ones were excluded. The MIDI velocity and MIDI volume information were translated to sound level using a calibration curve for the synthesizer used to produce the corresponding audio examples (Bresin and Friberg, 1997). However, the database contained few variations in sound level since many examples were coded using maximum values throughout the whole score.

*Pitch* (f0) was estimated by taking the average of all MIDI pitch values for each category.

*Articulation* (art) was computed as the average of the relative articulation for each note, given by the note duration divided by the inter-onset interval (IOI). IOI values longer than 800 ms were excluded to avoid outliers corresponding to rests in the track.

From previous studies (Madison and Paulin, 2010), it is evident that perceptual speed also depends on the perception of the beat. Therefore, the tempo specified in number of beats per second (BPS) was estimated manually by coauthor AE for each example.

The correlation between the computed MIDI features and perceptual features is shown in Table 4. Marked in bold are the correlations in which perceptual features supposedly have a direct relation with the computed features. As seen in the table, the strongest correlations occur for some of the features within these expected combinations with correlations in the range r = 0.7-0.8. For example, Speed is correlated with three features in this range (ann_tempo, nps_all, nps_dru). The main exception is for sound level, where none of the sound level features were found to be significantly correlated with perceptual Dynamics. The reason might be a weak variation in the database as discussed above, or that the sound level does not correspond to the perceived dynamics since it is expected to be coupled mainly to differences in timbre (Fabiani and Friberg, 2011). The remaining significant correlations are relatively small and in only one case exceed 0.6. They can presumably be due to either a skewness in the database (e.g. faster songs are played more staccato) or a covariance in the perceptual judgments (e.g. confusion between Pitch and Timbre).

*** Table 4 about here ***

The three different perceptual features directly corresponding to the MIDI features with significant correlations (Speed, Articulation, Pitch) were predicted using multiple regression.

In order to reduce the initial number of independent features, only a selection of the significant features in Table 4 was used for each perceptual feature. For Speed, for example, all the temporal features were included corresponding to the a priori important features. For the articulation and pitch features, only the overall features (art_all and f0_all) were included in the Speed regression since they were a priori considered to have a smaller effect.

In Table 5, the results for Speed are shown. The overall explained variance is about 90% in the multiple regression analysis. The significant features are only those that are intuitively directly related to temporal aspects. Note also that these features are all related to the accompaniment and mainly the drums while the melody is less important in this case. Due to the relatively large number of features and few songs, a partial least-square (PLS) regression was also performed (Geladi and Kowalski, 1986). PLS regression attempts to minimize the number of independent features by a principal component analysis in combination with a regression. The method can be used when there are a large number of inter-dependent features. The number of factors in the PLS regression was selected manually by choosing the minimum number that could still explain a major part of the variation. This was also cross-validated using 10 folds. With the modest number of three factors, this could still explain 85% of the variance using cross-validation. In this analysis, only the examples containing drums were included. Certainly, for other kinds of music such as classical music, the melodic parts often play a major role in defining the rhythm and thus also the perceived speed.

*** Table 5 about here ***

The prediction of Pitch is shown in Table 6. The result indicates a strong dependence on the melody since the pitch of the melody was the only significant contribution to the perceived Pitch. The overall explained variance (65%) was surprisingly small given that the pitch of each note is given in the MIDI representation. One possible reason could be the strong interaction between rated pitch and brightness (Table 2).

*** Table 6 about here ***

The prediction of Articulation is shown in Table 7. This indicates, as in the case of pitch, a rather straightforward relation, in particular to the melody (*sr*=0.51) and somewhat to the accompaniment (*sr*=0.16). Here the overall explained variance was 73%.

*** Table 7 about here ***

**VI. PREDICTION OF PERCEPTUAL FEATURES FROM AUDIO FEATURES**

**A. Computed features**

Due to the large number of suggested features in previous studies (e.g. Burred and Lerch, 2004) the approach here was to test the performance of existing audio features using available toolboxes. A number of low-level and mid-level audio features were extracted using the MIRToolbox (v. 1.3.1) (Lartillot and Toiviainen, 2007) and different VAMP plugins available in the Sonic Annotator[i] (v. 0.5). In total, 54 different features were computed (Table 8). Different feature sets were used for predicting the perceptual ratings. They were selected considering that they would *a priori* have a potential prediction influence. For example, Speed was expected to be influenced by different onset and tempo estimations. Such a selection could be found for six of the perceptual ratings, namely Speed, Rhythmic clarity, Dynamics, Modality, Timbre, and Brightness. Each feature was computed using the default settings and in certain cases using different available models. For each sound example, one final feature value was obtained. All onset measures were converted to onsets per second by counting the number of onsets and dividing by the total length of each music example. We found that the MIRToolbox yielded different results, particularly for high/mid-level features such as pulse clarity depending on the toolbox version. Thus, in order to replicate these results, the same version has to be used (v 1.3.1). For a more detailed description of the feature extraction for the ringtones in Exp. 1, see Hedblad (2011).

*** Table 8 about here ***

**B. Prediction of perceptual features from audio features**

An initial correlation analysis between ratings and audio features confirmed that many of the selected features correlated with the perceptual ratings. The size of the correlations was typically between 0.6 and 0.7 for features with a rather direct correspondence such as the pulse clarity computed in the MIRToolbox compared with the rated rhythmic clarity (r=0.73). The rated Speed correlated with different onset methods but surprisingly not with any tempo estimation at this stage. This is in contrast to the MIDI analysis above where we found a strong influence of annotated tempo, and is probably due to problems in correctly estimating the tempo of these examples.

A cross-correlation of the audio features showed, not surprisingly, quite high correlations in some cases. In particular, there were high correlations between similar features using different parameters, like different brightness measures. Most of these features were kept in the analysis. The exception was a few features in each experiment having a cross-correlation of 1.00, which were omitted in the prediction.

We used two different prediction methods: Partial Least-square Regression (PLS) and Support Vector Regression (SVR); see Table 9. Both were run in Matlab and for the SVR the LIBSVM package (Chang and Lin, 2011) was used. The SVR method was applied with a Gaussian radial bases kernel in most cases except for timbre and modality, in which a polynomial kernel was used. This was selected on a trial-and-error basis. As the number of features was relatively large compared to the number of songs, it was not meaningful to compute the direct fit of the models as in the analysis above. Thus, both methods were applied with a 10-fold cross-validation for the final estimation. The fit was estimated by the squared correlation coefficient $R^2$. In the application of the PLS, the number of independent components was selected manually for each prediction, optimizing the cross-validation result. Speed was analyzed using all onset and tempo features (14), Rhythmic clarity using the pulse clarity model (2), Modality using the modality features (3), while Dynamics, Timbre, and Brightness were analyzed using all spectral audio features. In addition, all available computational features were used as predictors of each perceptual feature (45-49).

*** Table 9 about here ***

The overall prediction was surprisingly low in many cases. For example, a large number of features related to onset and tempo prediction were computed from several toolboxes. Still the overall prediction of Speed for the ringtones in Exp. 1 obtained a modest value of about 70%. In comparison, the MIDI analysis obtained about 90% explained variance using only a few different onset measures combined with manual tempo estimation. This showed that Speed is related mainly to these features. Thus, the relatively poor performance in this case can be attributed to problems of extracting onsets and estimating tempo from audio. This is somewhat surprising considering both the relatively large previous research effort in this area, as well as the relatively low complexity of the ringtones since they were generated from MIDI and with often rather pronounced drums and clear onsets. This was the result of using default model parameters. We realize that the performance of different note detection methods, for example, would presumably improve significantly if the different model parameters were adjusted individually for these specific datasets.

The increased number of spectral features improved the prediction of Dynamics, which thus obtained the largest explained variance of 74% in Exp. 2. One would then be tempted to conclude that the spectral changes due to different dynamic level are reflected in this prediction. However, at the same time the prediction of the perceptual features timbre/brightness was surprisingly low.

**VII. CONCLUSIONS AND DISCUSSION**

The concept of using perceptually-derived music features was tested in a series of experiments using two different data sets. The result in terms of the inter-rater agreement, for example, was not consistent for all features but differed according to the type of feature and difficulty of the task. We can assume that if the feature is easily understood from a perceptual point-of-view it is also easy to rate. This implies that the traditional features naturally derived from a computational viewpoint are not necessarily the right choices. One obvious example is speed, which was both easy to rate, as indicated by the high agreement among the listeners, and possible to predict using different timing measures in the MIDI representation as well as from audio (Elowsson, Friberg, Madison and Paulin, 2013). Choosing the traditional features tempo or note density might have resulted in poorer ratings. We will now further discuss the research questions asked in the introduction.

*Do the raters agree in their judgments?* The agreement of the listeners varied considerably depending on the feature. The quality of a perceptual feature will always depend on what question we are asking and how clear it is for the subject to rate. One clear case of covariance seems to be the high correlation between Pitch and Brightness in the first experiment. Otherwise, both the relatively high subject inter-agreement and the rather modest feature cross-correlation indicate that the obtained averages of the ratings can be used to represent the perceptual ground truth of each feature.

*How many raters do we need?* This was not explicitly addressed in the current study but a high degree of both agreement and prediction using audio/MIDI features indicates that the number of raters is sufficient for these features. Also, prior to Exp. 1, a pilot experiment was run using 242 ringtones and a limited set of features. They were rated by five subjects who could be considered experts. The result of the pilot experiment (see Friberg, Schoonderwaldt and Hedblad, 2011) was not as consistent as Exp. 1 using similar examples but rated with 20 subjects. This and previous experience from other listening tests strongly indicates that it should be sufficient to use 20 subjects in order to obtain a reliable estimate of a perceptual feature of this kind. This was also supported in a recent experiment about emotion communication (Eerola, Friberg and Bresin, 2013). Two different groups (n=20, 26) rated the same examples using the same rating scales. The resulting mean values in this case were nearly identical, thus indicating that using more than about 20 subjects will not contribute much to the final estimate. For an efficient use of subjects in future experiments, it would be interesting to investigate this question further.

*Can higher-level semantic descriptions be accurately modeled from perceptual features?* The preliminary tests indicated an explained variance of about 90% in rather simple models using multiple regression. This is a first indication that this step in the analysis model (see Fig 1) has been simplified considerably. It is also an indication that the perceptual features represent or closely resemble some real perceptual phenomenon, as hypothesized in the first place. This was further supported by the agreement between the salient features in the multiple regression models and qualitative features used in previous research. The simplicity of such a model was also supported in a recent experiment by Eerola, Friberg and Bresin (2013). When a number of musical features were systematically varied, they found that the perception of the emotional expression was only to a very limited extent influenced by interaction effects between features and by nonlinearities (See also Juslin and Lindström, 2010).

*Can we make computational models of the perceptual features?* Several attempts at modeling the perceptual features using combinations of existing low- and mid-level computational features resulted in rather modest predictions of up to about 70% for the best cases. We believe, however, that this can be improved by developing models targeted specifically toward each feature. Recently, a model for speed using a limited set of custom rhythmic audio features has been developed (Elowsson, Friberg, Madison and Paulin, 2013). The results indicate that this model can perform at least as well as the MIDI model, explaining about 90% of the variation.

A major challenge in this project is to identify the perceptually relevant features. The use of speed instead of tempo or note density seems to be a relevant improvement toward a perceptually oriented description of overall features. The covariance between pitch and timbre found in these experiments point, on the other hand, toward an alternative description. This was also confirmed in an experiment by Schubert and Wolfe (2006) in which they found that perceptual brightness was influenced by the pitch. This connection was also observed informally by Alluri and Toiviainen (2009) when investigating polyphonic timbre. Interestingly, most computational measures related to timbre are also dependent on pitch, which is often pointed out as an issue in instrument detection, for example (Kitahara, Goto, Komatani, Ogata and Okuno, 2005). Thus, it might be plausible that in casual listening we use an overall combined perceptual feature consisting of both timbre and pitch. In fact, as suggested by Patterson, Gaudrain and Walters (2010), the traditional definition of pitch and timbre cannot explain how we perceive speech. In this case, the fundamental pitch as generated by the vocal folds together with the spectral envelope of the vocal tract determines the voice quality, i.e. the sex and the size of the speaker (Smith and Patterson, 2005), Thus, the ecological approach, which assumes that perception is about decoding the source ('who') and the message ('what') separately, seems to be a more fruitful methodological starting point.

Dynamics appeared to be rather easy to rate from the agreement between listeners. This feature also obtained rather good predictions using computed audio features. This indicates that dynamics is more consistently perceived and defined than timbre/brightness. This might be surprising considering that dynamics is a more complex parameter than brightness. It does, however, support the ecological hypothesis in which the primary goal of perception is to understand the source of the sound rather than the sound itself. In this case, the perceived dynamics is similar to perceiving the effort of the player(s).

In this study, the focus was on the prediction of emotional expression using perceptual features. However, we believe that the perceptual features selected in this study can be used for a general characterization of music and applied in other MIR tasks such as genre description. In this case, the features then need to be complemented with a further analysis of the sound sources such as a characterization of the instrumentation. The use of harmonic complexity was probably not an optimal choice as indicated by the low inter-rater correlation. This was a priori expected since it refers to a rather complex music theoretic analysis. However, it still was shown to be relevant for predicting the rated emotions of the film clips.

This indicates that a similar measure is needed that also includes some measure of inharmonicity or dissonance.

In summary, we think that the results so far are encouraging in that they indicate that perceptual features can be reliably estimated in listening experiments provided that there are at least 20 listeners rating the same examples. The modeling of these features is still a challenge, especially in some cases.

The "brute force" method using a variety of audio features for modeling the perceptual features was clearly not successful in these experiments. Certainly, more advanced machine learning methods and more features are likely to improve the results. However, we have shown that a careful selection of a limited set of features can be an efficient way of predicting both the perceptual features from MIDI, and higher-level semantic expressions. This indicates that the most important issue in music information retrieval tasks is the identification of appropriate and well-functioning features. Since the higher-level semantic prediction goals are perceptually based, it seems natural to base the features also on human perception. Potentially, the reduction of the problem into smaller subtasks as defined by perceptual features will lead to a clearer understanding of human perception and lead to better and simpler computational models.

**Acknowledgements**

This work was supported by the Swedish Research Council, Grant Nr. 2009-4285 and 2012-4685. Hamid Sarmadi developed the first version of the Matlab code for doing the SVR analysis and cross-validation.

**Footnotes**

[i] http://www.omras2.org/SonicAnnotator (date last viewed 9/1/14)

**References**

Alluri, V., and Toiviainen, P. (2009). "Exploring perceptual and acoustical correlates of polyphonic timbre," Music Perception, 27(3), 223-241.

Aucouturier, J.-J. and Bigand, E. (2013). "Seven problems that keep MIR from attracting the interest of cognition and neuroscience," J. Intell. Inf. Syst. 41, 483-497.

Aucouturier, J.-J. and Pachet. F. (2004). "Improving timbre similarity: How high is the sky? ," J. Negative Results in Speech and Audio Sciences, 1(1), 1-13.

Bresin, R., and Friberg, A. (1997). "A multimedia environment for interactive music performance," In Proc. KANSEI - The Technology of Emotion AIMI International Workshop, Genova, *1997*, pp. 64-67.

Burred, J. J. and Lerch, A. (2004). "Hierarchical Automatic Audio Signal Classification," J. Audio Eng. Soc., 52(7/8), 724-738.

Chang, C. C., and Lin, C. J. (2011). "LIBSVM: a library for support vector machines," ACM Trans. Intelligent Systems and Tech. (TIST), 2(3), 27, 1-39.

Clarke, E. F. (2005). *Ways of Listening: An Ecological Approach to the Perception of Musical Meaning* (Oxford University Press, Oxford, 240 pages).

Downie, J. S., West, K., Ehmann, A., and Vincent, E. (2005). "The 2005 music information retrieval evaluation exchange (mirex 2005): Preliminary overview," In proc. ISMIR 2005, 6th Int. Symp. on Music Information Retrieval, pp. 320-323.

Eerola, T. (2012). "Modeling Listeners' Emotional Response to Music," Topics in cognitive science, 4(4), 607-624.

Eerola, T. and Vuoskoski, J. K. (2011). "A comparison of the discrete and dimensional models of emotion in music," Psychol. of Music, 29 (1), 18-49.

Eerola, T., Friberg, A., and Bresin, R. (2013). "Emotional expression in music: contribution, linearity, and additivity of primary musical cues," Front. Psychol. 4(487), 1-12.

Elowsson, A., Friberg, A., Madison, G., and Paulin, J. (2013). "Modelling the Speed of Music Using Features from Harmonic/Percussive Separated Audio," In Proc. ISMIR 2013, Int. Symp. on Music Information Retrieval. (*no page numbers published*)

Fabiani, M., and Friberg, A. (2011). "Influence of pitch, loudness, and timbre on the perception of instrument dynamics," J. Acoust. Soc. Am. - Express Letters, EL193-EL199.

Friberg, A., and Ahlbäck, S. (2009). "Recognition of the main melody in a polyphonic symbolic score using perceptual knowledge," J. New Music Res. 38(2), 155-169.

Friberg, A., Bresin, R., and Sundberg, J. (2006). "Overview of the KTH rule system for musical performance," Advances in Cognitive Psychology, Special Issue on Music Performance, 2(2-3), 145-161.

Friberg, A. (2008). "Digital audio emotions — An overview of computer analysis and synthesis of emotions in music," In Proc. DAFx-08, the 11th Int. Conference on Digital Audio Effects, pp. 1-6.

Friberg, A., and Hedblad, A. (2011). "A Comparison of Perceptual Ratings and Computed Audio Features," In Proceedings SMC 2011, 8th Sound and Music Computing Conference, pp. 122-127.

Friberg, A., Schoonderwaldt, E., and Hedblad, A. (2011). "Perceptual ratings of musical parameters," In von Loesch, H., and Weinzierl, S. (Eds.), Gemessene Interpretation - Computergestützte Aufführungsanalyse im Kreuzverhör der Disziplinen (Klang und Begriff 4), (pp. 237-253), (Schott, Mainz).

Friberg, A., and Sundström, A. (2002). "Swing ratios and ensemble timing in jazz performance: Evidence for a common rhythmic pattern," Music Perception, 19(3), 333-349.

Gabrielsson, A., and Lindström, E. (2010). "The role of structure in the musical expression of emotions," In P. N. Juslin, and J. A. Sloboda (Eds.), Handbook of music and emotion: Theory, research, applications (pp. 367-400) (Oxford University Press, New York).

Gaver, W.W. (1993). "How Do We Hear in the World?: Explorations in Ecological Acoustics," Ecological Psychology, 5(4), 285-313.

Geladi, P., and Kowalski, B. R. (1986). "Partial least-squares regression: a tutorial," Analytica chimica acta, 185, 1-17.

Heckmann, M., Domont, X., Joublin, F., and Goerick, C. (2011). "A hierarchical framework for spectro-temporal feature extraction," Speech Communication, 53(5), 736-752.

Hedblad, A. (2011). *Evaluation of Musical Feature Extraction Tools Using Perceptual Ratings*. Master thesis, KTH Royal Institute of Technology, Stockholm; Sweden (35 pages).

Husain, G., Thompson, W. F., and Schellenberg, E. G. (2002). "Effects of musical tempo and mode on arousal, mood, and spatial abilities," Music Perception, 20(2), 151-171.

Ilie, G., and Thompson, W. F. (2006). "A comparison of acoustic cues in music and speech for three dimensions of affect," Music Perception, 23(4), 319-330.

ITU-R. (2006). *Rec. ITU-R BS.1770, Algorithms to measure audio programme loudness and true-peak audio level.* International Telecommunications Union.

Juslin, P.N., and Lindström, E. (2010). "Musical expression of emotions: Modelling listeners' judgements of composed and performed features," Music Analysis, 29(1-3), 334-364.

Ladefoged, P., and McKinney, N. P. (1963). "Loudness, Sound Pressure, and Subglottal Pressure in Speech," J. Acoust. Soc. Am. 35(4), 454-460.

Kitahara, T., Goto, M., Komatani, K., Ogata, T., and Okuno, H. G. (2005). "Instrument identification in polyphonic music: feature weighting with mixed sounds, pitch-dependent timbre modeling, and use of musical context," In Proc. ISMIR 2008, Int. Conf. on Music Information Retrieval, pp. 558-563.

Lartillot, O., and Toiviainen, P. (2007). "A MATLAB toolbox for musical feature extraction from audio," Proc. Of the 10$^{th}$ Int. Conference on Digital Audio Effects 2007 (DAFx-07), pp. 237-244.

Lartillot, O., Eerola, T., Toiviainen, P., and Fornari, F. (2008). "Multi-Feature Modeling of Pulse Clarity: Design, Validation and Optimization," In Proc. ISMIR 2008, Int. Conf. on Music Information Retrieval, pp. 521-526.

London, J. (2004). *Hearing in time: Psychological aspects of musical meter* (Oxford University Press, New York, 206 pages).

Lu, L., Liu, D., and Zhang H. (2006). "Automatic Mood Detection and Tracking of Music Audio Signals," IEEE Trans. on Audio, Speech, and Language Proc., 14 (1), 5-18.

Madison, G. and Paulin, J. (2010). "Ratings of speed in real music as a function of both original and manipulated beat tempo," J. Acoust. Soc. Am. 128(5), 3032-3040.

McGraw, K. O., and Wong, S. P. (1996). "Forming inferences about some intraclass correlation coefficients," Psychol. methods, 1(1), 30-46.

Mesgarani, N., Shamma, S., and Slaney, M. (2004). "Speech discrimination based on multiscale spectro-temporal modulations," In Proc. 2004 IEEE International Conference on Acoustics, Speech and Signal Processing, Vol. 1, pp. 601-4.

Nygren, P. (2009). *Achieving equal loudness between audio files.* Master thesis, KTH Royal Institute of Technology, Stockholm, Sweden (61 pages).

Pachet, F., and Roy, P. (2009). "Analytical features: a knowledge-based approach to audio feature generation," EURASIP J. Audio, Speech, and Music Processing, 2009, 1.

Patterson, R. D., Gaudrain, E., and Walters, T. C. (2010). "The perception of family and register in musical tones," In Music Perception (pp. 13-50), (Springer, New York).

Peeters, G. (2004). "A large set of audio features for sound description (similarity and classification) in the CUIDADO project," CUIDADO I.S.T. Project Report (25 pages).

Pohle, T., Pampalk, E., and Widmer, G. (2005). "Evaluation of frequently used audio features for classification of music into perceptual categories," In Proceedings of the International Workshop on Content-Based Multimedia Indexing. (*no page numbers found*)

Polotti, P. and Rocchesso, D. (Eds.) (2008). *Sound to Sense, Sense to Sound: A State of the Art in Sound and Music Computing,* (Logos Verlag, Berlin, 490 pages).

Rasch, R. A. (1979). "Synchronization in Performed Ensemble Music," Acustica, 43, 121-131.

Russell, J. A. (1980). "A circumplex model of affect," J. personality and social psychol., 39, 1161 - 1178.

Schubert, E., and Wolfe, J. (2006). "Does timbral brightness scale with frequency and spectral centroid?" Acta acustica united with acustica, 92(5), 820-825.

Slaney, M. (1998). "Auditory toolbox," Interval Research Corporation, Tech. Rep, 10 (52 pages).

Smith, D.R.R. and Patterson, R.D. (2005). "The interaction of glottal-pulse rate and vocal-tract length in judgements of speaker size, sex, and age," J. Acoust. Soc. Am. 125, 2374-2386

Wedin, L. (1972). "A Multidimensional Study of Perceptual-Emotional Qualities in Music," Scand. J. Psychol., 13, 241-257.

TABLE 1. Agreement among the subjects in Exp. 1 and 2 as indicated by mean inter-subject correlation and Cronbach's alpha. A value of 1 indicates perfect agreement in both cases. Cronbach's alpha above 0.9 is commonly considered an excellent fit.

|  | *Experiment 1 Ring tones* | | *Experiment 2 Film clips* | |
|---|---|---|---|---|
| *Feature* | corr. | alpha | corr. | Alpha |
| Speed | 0.71 | 0.98 | 0.60 | 0.97 |
| Rhy. complex. | 0.29 (0.33) | 0.89 (0.89) | 0.33 (0.34) | 0.91 (0.90) |
| Rhy. clarity | 0.31 (0.34) | 0.90 (0.90) | 0.48 (0.51) | 0.95 (0.95) |
| Articulation | 0.37 (0.41) | 0.93 (0.93) | 0.59 (0.60) | 0.97 (0.97) |
| Dynamics | 0.41 (0.44) | 0.93 (0.93) | 0.40 (0.49) | 0.93 (0.95) |
| Modality | 0.38 (0.47) | 0.93 (0.94) | 0.60 | 0.96 |
| Harm. complex. | 0.21 | 0.83 | 0.21 (0.25) | 0.85 (0.87) |
| Pitch | 0.37 (0.42) | 0.93 (0.93) | 0.43 (0.50) | 0.94 (0.94) |
| Brightness/Timbre | 0.27 | 0.88 | 0.32 (0.34) | 0.90 (0.91) |
| Energy | 0.57 | 0.96 | | |
| Valence | 0.42 (0.47) | 0.94 (0.94) | | |

TABLE 2. Cross-correlation of rated perceptual features in Exp. 1 and 2, * p<0.05; ** p<0.01, *** p<0.001.

Exp. 1 Ringtones, Number of songs = 100

|  | Speed | Rhy. Comp. | Rhy. Cla. | Articulation | Dynamics | Modality | Har. Comp. | Pitch |
|---|---|---|---|---|---|---|---|---|
| Rhy. Comp. | -0.13 | | | | | | | |
| Rhy. Cla. | 0.51*** | -0.56*** | | | | | | |
| Articulation | 0.56*** | -0.08 | 0.56*** | | | | | |
| Dynamics | 0.67*** | -0.04 | 0.56*** | 0.57*** | | | | |
| Modality | 0.21* | -0.16 | 0.02 | 0.21* | 0.04 | | | |
| Har. Comp. | -0.37*** | 0.52*** | -0.63*** | -0.49*** | -0.32** | -0.23* | | |
| Pitch | -0.03 | -0.04 | -0.18 | -0.08 | 0.03 | 0.45*** | 0.21* | |
| Brightness | 0.01 | -0.05 | -0.16 | -0.02 | 0.11 | 0.59*** | 0.15 | 0.90*** |

Exp. 2 Film clips, Number of songs = 110

|  | Speed | Rhy. Comp. | Rhy. Cla. | Articulation | Dynamics | Modality | Har. Comp. | Pitch |
|---|---|---|---|---|---|---|---|---|
| Rhy. Comp. | 0.30** | | | | | | | |
| Rhy. Cla. | 0.65*** | -0.21* | | | | | | |
| Articulation | 0.82*** | 0.31*** | 0.68*** | | | | | |
| Dynamics | 0.55*** | 0.21* | 0.34*** | 0.55*** | | | | |
| Modality | 0.40*** | -0.10 | 0.31*** | 0.32*** | 0.16 | | | |
| Har. Comp. | 0.15 | 0.62*** | -0.25** | 0.16 | 0.25** | -0.21* | | |
| Pitch | 0.14 | -0.03 | 0.07 | 0.07 | -0.00 | 0.40*** | 0.23* | |
| Timbre | 0.21* | -0.07 | 0.14 | 0.15 | 0.17 | 0.60*** | 0.13 | 0.90*** |

TABLE 3. Prediction of emotion features in Exp. 1 and 2.using linear regression. $R^2$ and *adj. $R^2$* reflects the total explained variance by each model, *sr* is the semipartial correlation coefficient reflecting the independent contribution of each feature. The minus sign indicates that the influence of that feature is negative. For clarity, only significant results are shown ($p < 0.05$), * p<0.05, ** p<0.01, *** p<0.001.

|  | *Experiment 1* |  | *Experiment 2* |  |  |  |  |  |  |  |
| --- | --- | --- | --- | --- | --- | --- | --- | --- | --- | --- |
|  | *Energy* | *Valence* | *Energy* | *Valence* | *Tension* | *Anger* | *Fear* | *Happin.* | *Sadness* | *Tenderness* |
| $R^2$ | 0.94 | 0.90 | 0.92 | 0.80 | 0.80 | 0.74 | 0.67 | 0.83 | 0.77 | 0.65 |
| *Adjusted $R^2$* | 0.93 | 0.88 | 0.91 | 0.78 | 0.79 | 0.72 | 0.64 | 0.81 | 0.75 | 0.62 |
| *Cross val. $R^2$* | 0.93 | 0.88 | 0.90 | 0.75 | 0.76 | 0.68 | 0.59 | 0.78 | 0.72 | 0.58 |
| **Feature** | *sr* |  | *sr* |  |  |  |  |  |  |  |
| Speed | 0.36*** | 0.09* | 0.14*** | 0.10* |  |  |  | 0.10* |  |  |
| Rhy.complex. |  |  |  |  |  |  |  |  |  |  |
| Rhy.clarity | 0.08** |  |  | 0.10* |  |  | (-)0.13* |  |  |  |
| Articulation |  | 0.07* | 0.11*** | (-)0.10* | 0.15** |  | 0.17** |  | 0.18*** | (-)0.18** |
| Dynamics | 0.20*** | (-)0.13*** | 0.39*** | (-)0.27*** | 0.37*** | 0.50*** | 0.25*** |  | (-)0.18*** | (-)0.37*** |
| Modality | 0.10*** | 0.49*** | 0.13*** | 0.27*** | (-)0.18*** |  |  | 0.37*** | (-)0.44*** | 0.17** |
| Harm. complex. |  |  |  | (-)0.21*** | 0.21*** | 0.17** | 0.30*** | 0.10* | (-)0.22*** |  |
| Pitch |  |  | 0.07* |  |  |  |  |  |  |  |
| Brightness Timbre |  | 0.10** | (-)0.12*** | 0.15** | (-)0.16*** | (-)0.15** | (-)0.23*** |  | 0.11* |  |

TABLE 4. Correlation between computed MIDI features and perceptual features in Exp. 1. There were 100 songs in total but for each parameter there were some missing values. Only significant results are shown (p<0.05), * p<0.05; ** p<0.01, *** p<0.001.

| Feature | Speed | Rhythmic complexity | Rhythmic clarity | Articulation | Dynamics | Modality | Harmonic complexity | Pitch | Brightness |
|---|---|---|---|---|---|---|---|---|---|
| ann_tempo | **0.77**\*** | -0.23* | 0.53*** | 0.47*** | 0.47*** | | -0.34*** | | |
| nps_all | **0.72**\*** | | 0.35*** | 0.44*** | 0.51*** | | -0.22* | | |
| nps_mel | **0.40**\*** | | | 0.36*** | 0.30** | | -0.23* | | |
| nps_acc | **0.41**\*** | | 0.22* | 0.28** | 0.34** | | | | |
| nps_bas | **0.65**\*** | -0.23* | 0.38*** | | 0.40*** | | -0.24* | | |
| nps_dru | **0.71**\*** | | 0.26* | 0.51*** | 0.60*** | | -0.26* | | |
| nps_dru_tom | **0.55**\*** | | 0.27* | 0.39*** | 0.46*** | 0.27** | -0.26* | | |
| nps_dru_rest | **0.55**\*** | | | 0.32** | 0.44*** | | | | |
| sl_all | | | 0.28** | | | | | | |
| sl_mel | | | | | | | | | |
| sl_acc | | | | | | | | | |
| sl_bas | | | 0.26* | | | | | -0.23* | |
| sl_dru | | | 0.33** | | | | | | |
| f0_all | -0.27** | | -0.31** | | | | 0.32** | **0.68**\*** | 0.58*** |
| f0_mel | | | | | | 0.20* | 0.22* | **0.80**\*** | 0.65*** |
| f0_acc | -0.23* | | | | | | | **0.29**\** | 0.31** |
| f0_bas | | | | | | | 0.25* | **0.25**\* | |
| art_all | -0.47*** | | -0.45*** | **-0.72**\*** | -0.36*** | -0.20* | 0.40*** | | |
| art_mel | -0.36*** | | -0.30** | **-0.83**\*** | -0.33*** | | 0.38*** | | |
| art_acc | -0.36*** | | -0.39*** | **-0.50**\*** | -0.30** | -0.23* | 0.28** | | |
| art_bas | -0.35*** | | -0.46*** | **-0.51**\*** | -0.23* | | 0.32** | | |

TABLE 5. Speed predicted from MIDI features using multiple regression and PLS regression. $R^2$ is the overall correlation or explained variance, adjusted $R^2$ with compensation for the number of prediction variables, $β$ is the beta coefficient, *sr* the semi-partial correlation coefficient, and *p* is the probability of a significant contribution from each feature, * p<0.05; ** p<0.01, *** p<0.001.

| *Number of songs* | 66 | | |
|---|---|---|---|
| *Multiple regression* | | *PLS regression* | |
| $R^2$ | 0.91 | *factors* | 3 |
| *Adjusted $R^2$* | 0.89 | *Adj $R^2$* | 0.90 |
| | | *Adj $R^2$ cross-val.* | 0.85 |
| *Multiple regression* | | | |
| *feature* | *β* | *sr* | *p* |
| ann_tempo | 0.452 | 0.312 | 0.000*** |
| nps_mel | 0.046 | 0.039 | 0.350 |
| nps_acc | 0.141 | 0.129 | 0.003** |
| nps_bas | 0.139 | 0.088 | 0.037* |
| nps_dru | 0.058 | 0.028 | 0.495 |
| nps_dru_tom | 0.194 | 0.123 | 0.004** |
| nps_dru_rest | 0.303 | 0.175 | 0.000*** |
| f0_all | -0.082 | 0.062 | 0.141 |
| art_all | -0.014 | 0.011 | 0.784 |

TABLE 6. Pitch predicted from MIDI features using multiple regression. For explanation of terms, see Table 5.

| Number of songs | 79 |
|---|---|
| $R^2$ | 0.67 |
| Adjusted $R^2$ | 0.65 |

| feature | β | sr | p |
|---|---|---|---|
| sl_bas | -0.123 | 0.121 | 0.073 |
| f0_mel | 0.776 | 0.726 | 0.000*** |
| f0_acc | 0.090 | 0.088 | 0.192 |
| f0_bas | 0.002 | 0.002 | 0.978 |

TABLE 7. Articulation predicted from MIDI features using multiple regression. For explanation of terms, see Table 5.

| Number of songs | 76 |
|---|---|
| $R^2$ | 0.74 |
| Adjusted $R^2$ | 0.73 |

| feature | β | sr | p |
|---|---|---|---|
| nps_dru | 0.137 | 0.118 | 0.054 |
| art_mel | -0.658 | 0.512 | 0.000*** |
| art_acc | -0.200 | 0.163 | 0.008** |
| art_bas | -0.048 | 0.037 | 0.535 |

TABLE 8. Overview of the calculated audio features, * Exp. 1 only, ** Exp. 2 only.

| Feature | Abbreviation | Number of features | Comment |
|---|---|---|---|
| **MIR Toolbox** | | | |
| Zero crossings | MT_ZCR | 1 | |
| MFCC | MFCC | 13 | |
| Brightness | MT_Bright | 3 | Cutoff : 1, 1.5, 3 kHz |
| Spectral centroid | MT_SC | 1 | |
| Spectral shape | S_Spread, S_Skew, S_Kurtos, S_Flat | 4 | Spread, skewness, kurtosis, flatness |
| Spectral roll off | SRO | 2 | 85,95 |
| Spectral flux | MT_Flux | 1 | |
| Attack time | MT_Att time | 2 | |
| Attack slope | MT_Att slope | 2 | |
| RMS | RMS | 1 | |
| Silence ratio | MT_ASR | 1 | |
| Event density | MT_Event | 1 | |
| Pulse clarity | MT_Pulse_clarity | 2 | Model: 1,2 |
| Tempo | MT_Tempo | 3 | Model: autocorr, spectrum, both |
| Mode | MT_Mode | 2 | Model: best, sum |
| **aubio** | | | |
| Onset | aubio_onset | 2** | |
| **EX** | | | |
| Onset | ex_onsets | 1* | |
| Tempo | ex_tempo | 1 | |
| **Mazurka** | | | |
| Beat | mz_beat | 1 | |
| Onset | mz_sf_onset, mz_sf_onset, mz_srf_onset | 3** | Flux, reflux |
| **Queen Mary** | | | |
| Mode | qm_mode | 1 | |
| Onset | qm_onsets_f , qm_onsets_s | 2 | |
| Tempo | qm_tempo | 1 | |
| **DS** | | | |
| Onset | ds_onsets_f, ds_onsets_s | 2* | |
| | | | |
| Pitch | EN_F0_Key_S | 1** | Custom average pitch |

TABLE 9. The prediction of perceptual features from audio features in Exp. 1 and 2 in terms of the squared correlation coefficient $R^2$ using Partial Least-square Regression (PLS) and Support Vector Regression (SVR) with 10-fold cross validation.

|  |  | *Experiment 1 Ringtones* | | | | *Experiment 2 Film clips* | | | |
|---|---|---|---|---|---|---|---|---|---|
| *Perceptual Feature* | *Selection* | *# of features* | *PLS comp.* | *$R^2$ PLS* | *$R^2$ SVR* | *Audio features* | *PLS comp.* | *$R^2$ PLS* | *$R^2$ SVR* |
| Speed | timing | 13 | 3 | 0.69 | 0.74 | 16 | 2 | 0.37 | 0.51 |
|  | all | 45 | 2 | 0.63 | 0.63 | 49 | 2 | 0.31 | 0.56 |
| Rhythmic compl. | timing | 13 | 2 | 0.10 | 0.19 | 16 | 2 | 0.0 | 0.04 |
|  | all | 45 | 2 | 0.03 | 0.06 | 49 | 2 | 0.0 | 0.07 |
| Rhythmic clarity | pulse clarity | 2 | 1 | 0.49 | 0.37 | 2 | 1 | 0.23 | 0.29 |
|  | timing | 13 | 3 | 0.51 | 0.41 | 16 | 2 | 0.32 | 0.42 |
|  | all | 45 | 1 | 0.32 | 0.32 | 49 | 2 | 0.31 | 0.42 |
| Articulation | all | 45 | 2 | 0.49 | 0.43 | 49 | 2 | 0.39 | 0.47 |
| Dynamics | timbre | 25 | 3 | 0.48 | 0.58 | 25 | 3 | 0.60 | 0.74 |
|  | all | 45 | 3 | 0.53 | 0.56 | 49 | 3 | 0.54 | 0.66 |
| Modality | modality | 3 | 2 | 0.43 | 0.40 | 3 | 2 | 0.47 | 0.47 |
|  | all | 45 | 2 | 0.38 | 0.32 | 49 | 4 | 0.53 | 0.47 |
| Pitch | all | 45 | 2 | 0.26 | 0.31 | 49 | 2 | 0.31 | 0.27 |
| Timbre | timbre |  |  |  |  | 25 | 4 | 0.37 | 0.33 |
|  | all |  |  |  |  | 49 | 4 | 0.41 | 0.37 |
| Brightness | timbre | 25 | 2 | 0.11 | 0.04 |  |  |  |  |
|  | all | 45 | 2 | 0.31 | 0.21 |  |  |  |  |

FIGURE CAPTION

FIG. 1. The different layers of audio analysis in the suggested method using perceptual features.

# Semantic description
**Genre, emotion, motional qualities …**

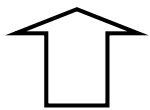

# Perceptual features
**mode, harmonic complexity, speed, rythmic clarity…**

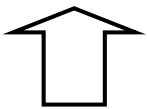

# Low-level Audio features
**sound level, MFCC, event rate …**

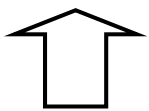

# Audio input

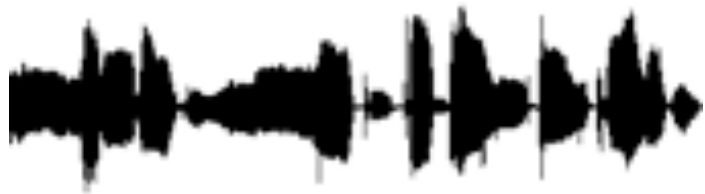